\documentclass[useAMS,usenatbib,A4]{mn2e}
\usepackage{graphicx}% This allows you to call \includegraphics
\def\etal{{\it et al}}
\def\deg{^{\circ}}
\def\P3hat{{\mathaccent 94 P}_3}

\def\eg{{\it e.g.}}
\def\ie{{\it i.e.}}
\def\cf{{\it cf.}}
\def\aap{A\&A}
\def\aaps{A\&A Suppl.}
\def\apj{Ap.J.}
\def\apjs{Ap.J. Suppl.}
\def\mnras{M.N.R.A.S.}
\def\nat{Nature}
\def\apss{Ap\&SS}
\newcount\notenumber
\def\clearnotenumber{\notenumber=0}
\def\note{\advance\notenumber by1 \footnote{$^{\the\notenumber}$}}
\clearnotenumber
\usepackage{rotating}   % to rotate the expressions and tables.
\title[PSR B0329+54:  A Glimpse of the Core Emission Process]
{Absolute Broadband Polarization Behaviour of  PSR B0329+54:  A Glimpse of the Core Emission Process}

\author[Dipanjan Mitra, Joanna Rankin \& Yashwant Gupta] 
{Dipanjan Mitra$^{1}$, Joanna M. Rankin$^{2}$ \& Yashwant Gupta$^{1}$ \\ 
$^1$National Centre for Radio Astrophysics, Ganeshkhind, Pune 411 007 India : dmitra@ncra.tifr.res.in; ygupta@ncra.tifr.res.in \\
$^2$Physics Department, University of Vermont, Burlington, VT 05405 USA : Joanna.Rankin@uvm.edu \\
}

\date{Released 2004 Xxxxx XX}

\pagerange{\pageref{firstpage}--\pageref{lastpage}} \pubyear{2004}

\def\LaTeX{L\kern-.36em\raise.3ex\hbox{a}\kern-.15em
    T\kern-.1667em\lower.7ex\hbox{E}\kern-.125emX}

\begin{document}

\label{firstpage}

\maketitle

\begin{abstract}
In this paper we report multifrequency single pulse polarization observations 
of the PSR B0329+54 normal mode using the Giant Meterwave Radio Telescope 
at 325 and 610 MHz and the Effelsberg Observatory at 2695 MHz. Our observations 
show that towards the central part of the polarization position-angle traverse there 
is an unusual ``arc''-like structure, which is comprised of a broadband ``kink'' and 
a frequency-dependent ``spot.''  The features are intimately connected with the 
intensity dependence of the core component:  the stronger emission arrives 
earlier and its linear polarization is displaced farther along the ``kink''.  Moreover, 
at high intensities, the circular polarization is --/+ antisymmetric; the nearly complete 
positive circular is characteristic of the weaker, later core subpulses.  We 
find that the ``kink'' emission is associated with the extraordinary (X) propagation 
mode, and hence propagation effects do not appear capable of producing the core 
component's broadband, intensity-dependent emission.  Rather, the overall 
evidence points to a largely geometric interpretation in which the ``kink'' provides 
a rare glimpse of the accelerating cascade or height-dependent amplifier 
responsible for the core radiation.  
\end{abstract}

\begin{keywords}
 miscellaneous -- methods:MHD --- plasmas --- data analysis --  s: general, individual (B0329+54) --- radiation mechanism: nonthermal -- polarization.
\end{keywords}

\section*{I. Introduction} 
\label{sec:intro}
Radio pulsars are well known to emit highly polarized radiation. 
Radhakrishnan \& Cooke (1969; hereafter R\&C) observed the Vela 
pulsar and demonstrated that the linear polarization position angle 
(PA) across its average-pulse profile exhibits a characteristic 
`S-shaped' traverse. This PA-longitude dependence was interpreted 
as reflecting the varying projected magnetic field direction within an 
overall dipolar configuration (R\&C; Komesaroff 1970). According to 
this rotating-vector model (RVM), wherein the received emission is 
associated with those field lines (momentarily) having tangents along 
our sight-line direction, the linear polarisation will be oriented within 
this tangent plane, varying with the neutron star's rotation.  The 
PA $\chi$ as a function of pulse longitude $\varphi$ can then be 
represented as
\begin{equation}
\chi=\tan^{-1}\left(\frac{\sin\alpha \sin\varphi}{\sin\xi
 \cos\alpha-\sin\alpha \cos\xi\cos\varphi}\right).
\label{eq1}
\end{equation}
(Manchester \& Taylor 1977), where $\alpha$ is the magnetic latitude 
and $\beta$ the sightline impact angle such that $\xi =\alpha + \beta$. 

The RVM strongly suggests that pulsar radiation is highly beamed 
and results from relativistic charged particles moving along the open 
dipolar magnetic field lines producing curvature radiation. This simple 
RVM model, however has had difficulty in explaining the wide diversity 
of average PA  traverses (\eg, Gould \& Lyne 1998; Hankins \& Rankin 
2006) observed in pulsar average profiles. Most pulsars, for instance, 
exhibit orthogonal polarization modes (OPMs) wherein the PAs are 
found to have two preferred values at a given pulse longitude differing 
by about 90\degr\ (Manchester \etal\ 1975; Backer \& Rankin 1980). 
The relative strengths of the modes vary with longitude causing the 
average PA traverse to exhibit ``90\degr\  jumps.''  Even when these 
modal effects are carefully considered, many pulsars show average 
PA behaviours that are inconsistent with the RVM (Everett \& Weisberg 
2001; Ramachandran \etal\ 2004).  Individual pulse studies, however, 
more clearly delineate OPM effects, and the behaviour of each mode 
is more usually found to be consistent with the RVM.  

Pulsar B0329+54 provides a fascinating context for investigating 
detailed polarization behaviour.  Discovered in 1968 and reported in 
the third and final Cambridge ``batch'' (Cole \& Pilkington 1968), it is 
one of the brightest pulsars in the northern sky and has thus been 
studied both early and very extensively, often with the Effelsberg 
100-m telescope at 1700 MHz.  One such classic study by Bartel 
\etal\ (1982; hereafter BMSH) carefully delineated the properties of 
its profile and polarisation modes.  A more recent observation by 
Gil \& Lyne (1995; hereafter GL95) using the Lovell instrument at 
408 MHz exhibited the remarkable complexity of the object's PA 
behaviour and  stimulated Mitra (1999) to investigate the physical 
origins of its non-RVM effects.  Referring to GL95's fig. 1, we see 
that the pulsar's average PA is distorted by the presence of the two 
OPMs over much of its duration.  The individual pulse PAs are well 
enough confined to the two modal ``tracks'', though, that consistent, 
reliable fits to eq.(1) could be made to them.  These authors do also 
note,  however, that the emission shows significant departures from 
the RVM within a region near the center of the profile associated with 
its core component---a fact noted earlier as well using modal profiles 
by Gil \etal\ (1992).  

Peculiar PA behaviour has been noted in a number of other pulsars 
within the longitude range of the central core component (\eg, Rankin 
1983a, 1990; hereafter R83a, R90). In the core-cone emission 
model, the central core and outlying conal components are thought to 
be produced by a central ``pencil'' beam within a hollow radiation cone.
In such configurations (of which B0329+54 is an excellent example), 
the RVM behaviour is most clearly associated with the conal components 
and is often interrupted or distorted in the core region.  This apparent 
non-RVM emission has even prompted speculation that the core might 
be produced by a different emission mechanism than that of the cone
(\eg, Radhakrishnan \& Rankin 1990), but this conjecture has attracted 
no satisfactory theoretical grounding.  

More recently, Malov \& Suleymanova (1998: hereafter MS98) as well 
as Gangadhara \& Gupta (2001; hereafter G\&G) have attempted to 
interpret B0329+54's emission configuration on the basis of aberration 
and retardation (hereafter A/R), taking its bright central feature as the 
core component---and therefore as the profile center---and then computing 
emission heights for its several cones.  The latter authors also find 
evidence for several new emission components, and we will use 
their designations below (see their fig. 3).  And in two other papers 
Suleymanova \& Pugachev (1998, 2002; SP98/02) first analyse the 
linear polarisation distributions at 103 and 60 MHz and then study 
transitions between the pulsar's ``abnormal'' and ``normal'' profile 
modes at 111 MHz.  Karastergiou \etal\  (2001) study how power and 
polarisation are correlated in simultaneous observations at 1.41 and 
2.69 GHz.  

Edwards \& Stappers (2004; hereafter ES04) have published a major 
study of the pulsar's OPM properties using new high quality 328-MHz 
polarimetric observations from the Westerbork Synthesis Radio Telescope 
(WSRT).  Their novel analysis delineates the character of the pulsar's 
OPM behaviour in detail, and on this basis they speculate about 
its physical origins.  Their study underscores the importance of 
understanding the non-RVM, core-component properties in 
the full context of both current analyses and the rich published 
literature.  We will make considerable use of ES04's results in our work 
below.  

Currently, a number of divergent ideas have arisen in the course of 
efforts to understand OPM anomalies:  GL95 explained their results 
as an effect of finite beamwidth wherein radiation from nearby field 
lines was superposed within the sightline.  Further, they argued that 
this effect is more severe near the polar cap from where the field 
lines diverge. (In general, however, radiation from nearby field lines 
should superpose symmetrically and should not affect the overall PA 
traverse; a very special circumstance is needed to produce the
observed non-RVM behaviour.) Mitra \etal\ (2000) attributed this 
effect to multipolar magnetic field contributions in the core-emission 
region.  More recently Mitra \& Sieradakis (2004) have speculated that 
A/R resulting from height-dependent emission can cause distorted PA 
traverses, and ES04 appeal to magnetospheric refraction to the same 
end.  Srostlik \& Rankin (2005) have shown that OPM's associated 
with the core emission in B1237+25 can distort the average PA traverse.
Ramachandran \etal\ (2004) attributed PA anomalies in B2016+28 to 
return currents in the pulsar magnetosphere. Given the multitude of 
possible explanations available, it is clear that primary new observational 
and analytical work on the OPM phenomenon is needed, especially 
in the core-emission context.  Some important constraints, for instance, 
can be obtained by studying the frequency dependence of the OPM 
phenomena responsible for the non-RVM PA behaviours.

Furthermore, B0329+54's bright central component exhibits a clear 
intensity dependence:  At low intensities its peak lies most of a degree 
later than at high intensities as shown by McKinnon \& Hankins' 
(1993; hereafter MH93) fig. 1.  They also find that the overall profile 
width decreases with increasing intensity.  Such behaviour is unusual, 
although the Vela pulsar B0833--45 is also known to exhibit a similar 
effect (Krishnamohan \& Downs 1983; hereafter KD83).  Such an 
overall shift in the central component position is also seen between 
the pulsar's two profile modes, such that the abnormal lags the normal 
by some 0.5\degr. (see BMSH: fig. 3 or SP02: fig. 1).  We will see below 
that B0329+54's non-RVM effects---like those in Vela above---are
closely associated with the intensity-dependent position of its central 
component---so we will take KD's classic analysis and interpretation 
as a starting point for our work below.

\begin{figure} 
\begin{center}
\includegraphics[width=78mm,angle=-90.]{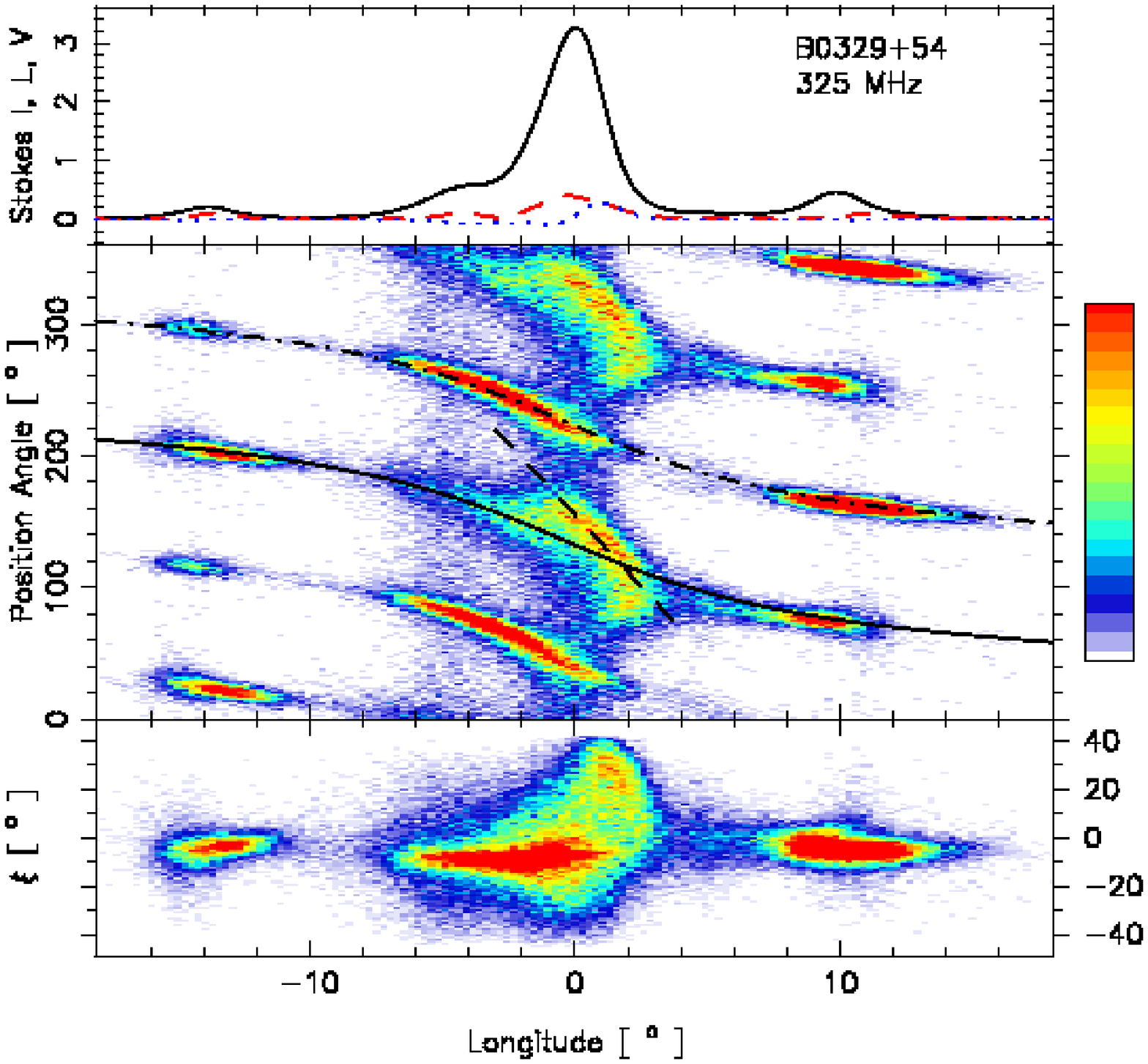}
\includegraphics[width=78mm,angle=-90.]{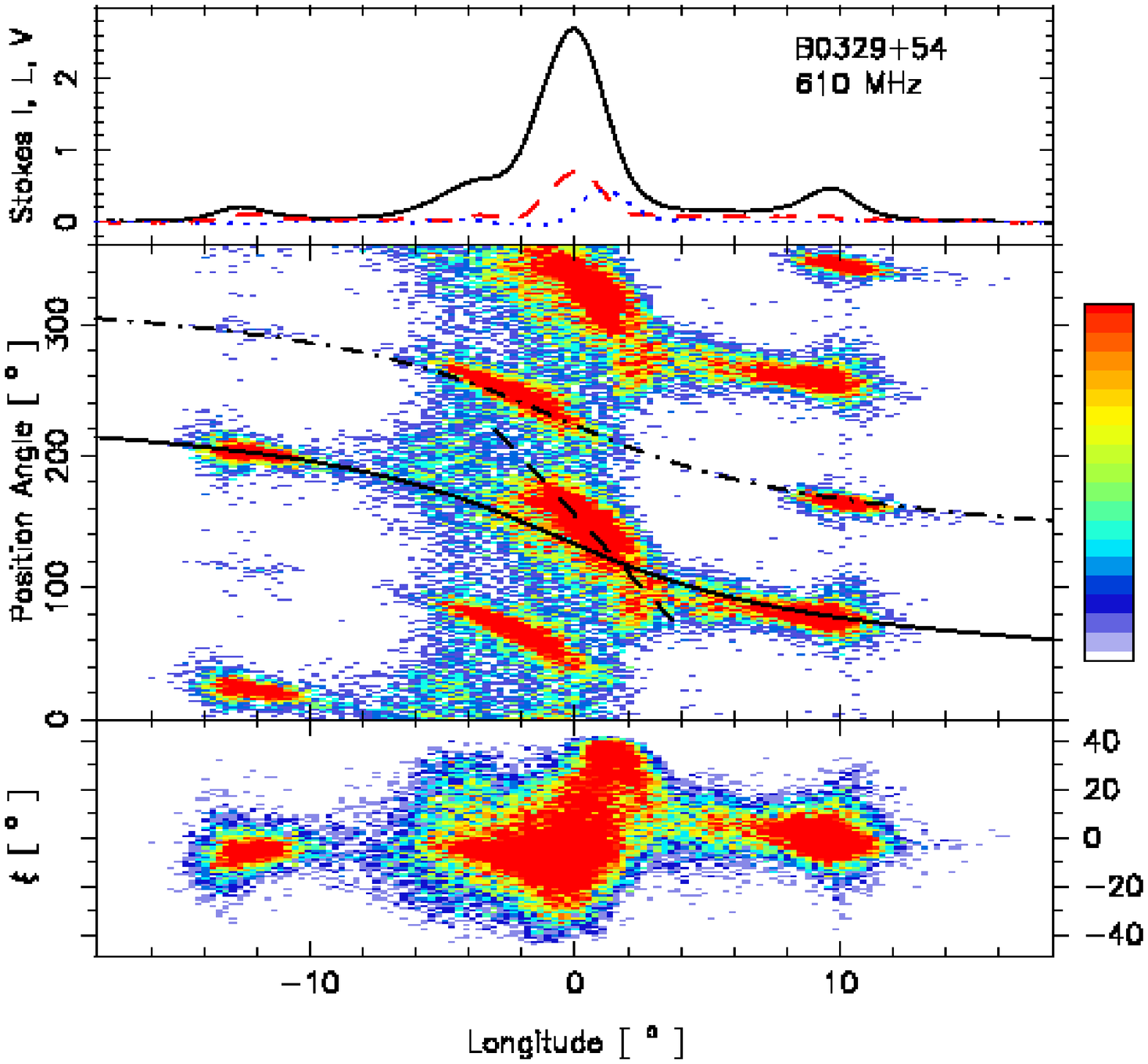}
\end{center}
\end{figure}

\begin{figure} 
\begin{center}
\includegraphics[width=78mm,angle=-90.]{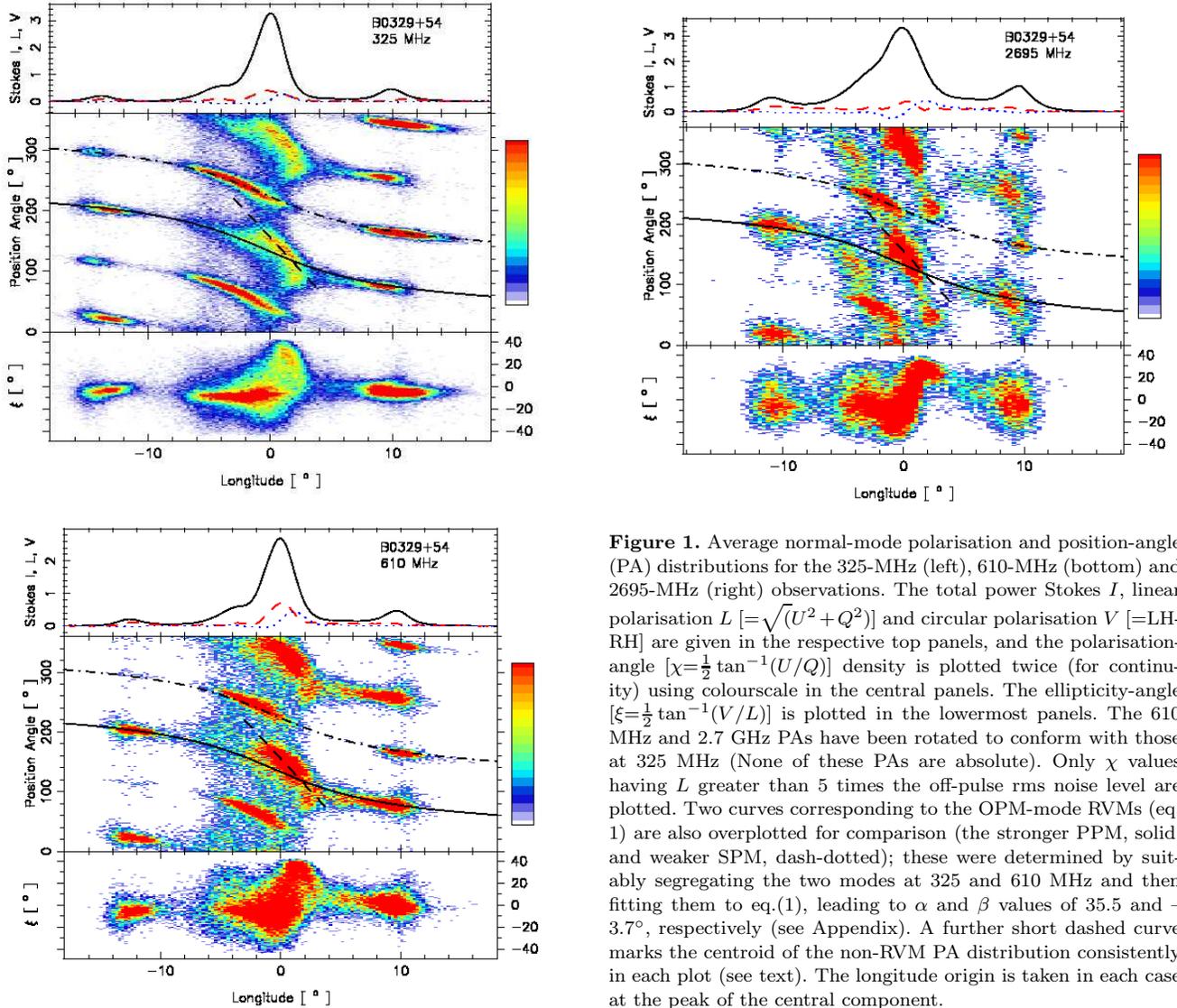}
\caption{Average normal-mode polarisation and position-angle (PA) 
distributions for the 325-MHz (left), 610-MHz (bottom) and 2695-MHz 
(right) observations.  The total power Stokes $I$, linear polarisation 
$L$ [=$\sqrt(U^2 + Q^2)$] and circular polarisation $V$ [=LH-RH] are 
given in the respective top panels, and the polarisation-angle  
[$\chi$=${1 \over 2}\tan^{-1}$($U$/$Q$)] density is plotted twice (for 
continuity) using  colourscale in the central panels.
The ellipticity-angle [$\xi$=${1 \over 2}\tan^{-1}$($V$/$L$)] 
is plotted in the lowermost panels. The 610 MHz and
2.7 GHz PAs have been rotated to conform with those at 325 MHz 
(None of these PAs are absolute).  Only $\chi$ values having $L$ 
greater than 5 times the off-pulse rms noise level are plotted.  Two 
curves corresponding to the OPM-mode RVMs (eq. 1) are also 
overplotted for comparison (the stronger PPM, solid, and weaker 
SPM, dash-dotted); these were determined by suitably segregating 
the two modes at 325 and 610 MHz and then fitting them to eq.(1), 
leading to $\alpha$ and $\beta$ values of 35.5 and --3.7$\deg$, 
respectively (see Appendix).  A further short dashed curve marks the 
centroid of the non-RVM PA distribution consistently in each plot 
(see text).  The longitude origin is taken in each case at the peak of 
the central component.}
\label{fig1}
\end{center}
\end{figure}

\begin{table}
\label{tab1}
 \centering
 % \begin{minipage}{140mm}
   \caption{Single-Pulse Polarimetry Observations}
   \begin{tabular}{ccccc}
   \hline
     Telescope  &    &  BW & Resolution & Pulses  \\
     Frequency  &  Date   &  (MHz)  &(msec) & (\#) \\
       (MHz)    &         & channels  &  (\degr)    &   Mode \\
   \hline
   GMRT  & 2004  &  16   & 0.512 & 2970 \\
   325  & 27 Aug  &  256  &  0.26  & normal  \\
   GMRT  & 2005     & 0.16  & 0.512 & 1650 \\
   610  & 1 Jun     & 128  &  0.26  & normal \\
   Effelsberg & 1997 &  80   & 0.697 & 2267 \\
   2650 & 19 Oct & 128   &  0.35  & normal \\
 \hline
 \end{tabular}
 %\end{minipage}
 \end{table}

In this paper we revisit the issues encountered in understanding the 
PA distributions of pulsar B0329+54 using high quality polarimetry 
spanning some three octaves.  In \S II we give details of the observations 
and their preparation for analysis. \S III discusses the multifrequency 
PA distributions, and \S IV considers means to segregate the RVM and 
non-RVM behaviours.  In \S V we discuss the several distinct contributions 
to core power,  and \S VI provides a discussion and review of the results.     
Two short appendices then review evidence pertaining to PSR B0329+54's 
basic conal and core emission geometry as well as efforts to confirm and 
extend it. 

%Outline
%I. Introduction

%II. Observations

%III. Multifrequency PA Distributions

%IV. Segregating the RVM and non-RVM Behaviours

%V. Geometrical issues

%?. Discussion

\section*{II. Observations} 
\label{sec2:II}
Pulse-sequence (hereafter PS) polarisation observations of pulsar 
B0329+54's `normal' mode at 325, 610 and 2650 MHz are described 
in Table 1 and presented below. The 325- and 610-MHz PSs were 
acquired using the Giant Meterwave Radio Telescope (GMRT) near 
Pune, India and the 2695-MHz observations were made with the 
100-m Effelsberg telescope, near Bonn, Germany.  The GMRT is a 
multi-element aperture-synthesis telescope (Swarup \etal\ 1991) 
consisting of 30 antennas distributed over a 25-km diameter area 
which can be configured as a single dish both in coherent and 
incoherent array modes of operation.   The polarimetric observations 
discussed here used the coherent (or more commonly called  
`phased array') mode (Gupta \etal\ 2000; Sirothia 2000) in the 
upper of the two 16-MHz `sidebands'.  At either frequency right- 
and left-hand circularly polarized complex voltages arrive at the 
sampler from each antenna. The voltage signals are subsequently 
sampled at the Nyquist rate and processed through a digital receiver 
system consisting of a correlator, the GMRT array combiner (GAC), 
and a pulsar back-end.  In the GAC the signals selected by the user 
are added in phase and fed to the pulsar backend. The pulsar 
back-end computes both the auto- and cross-polarized power levels, 
which were then recorded at a sampling interval of 0.512 msec.  A 
suitable calibration procedure as described in Mitra \etal\ (2005) is 
applied to the recorded data to get the calibrated Stokes parameters 
$I$, $Q$, $U$ and $V$.  The PS at 2695 MHz is an archival Effelsberg 
observation.  The polarimetry there was carried out with a multiplying 
polarimeter and calibrated using the procedure described by von 
Hoensbroech \& Xilouris (1997) and von Hoensbroech (1999).  The 
calibrated Stokes PSs at all the frequencies were finally converted into 
European Pulsar Network (EPN, Lorimer \etal\ 1998) format for offline 
analysis.

\begin{figure*} 
\begin{center}
\includegraphics[width=160mm]{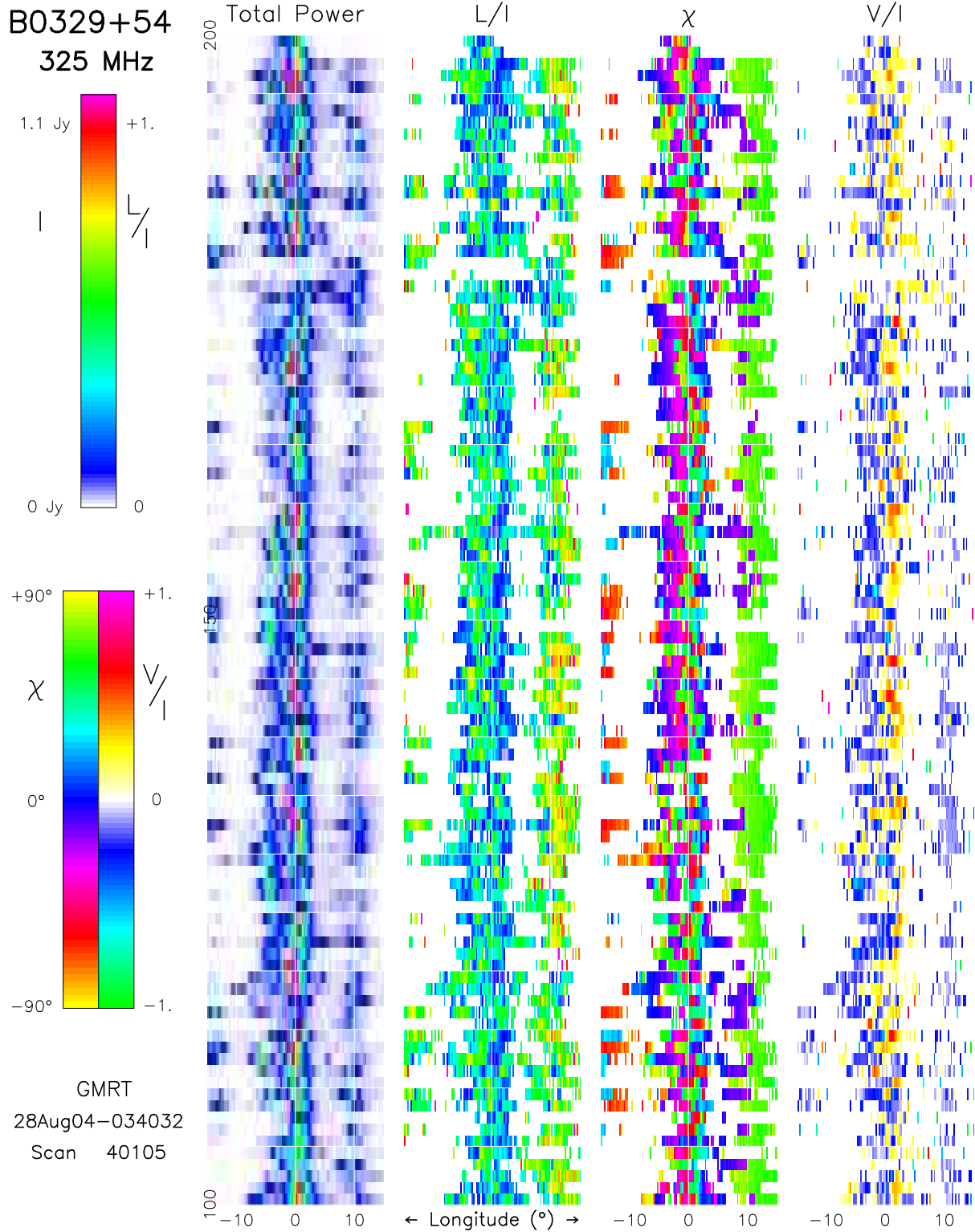}
\caption{Pulse-sequence polarization display showing a 100-pulse 
section of the 325-MHz observations in Fig.~\ref{fig1} (top left).  The 
(uncalibrated) total power $I$, fractional linear $L/I$, PA $\chi$, and 
fractional circular polarization $V/I$ are colour-coded in each of four 
columns according to their respective scales at the left of the diagram.  
Note that the intensity of the central core component varies strongly, 
from being undetectable during ``core nulls'' (\eg, pulses 180--181) to 
nearly saturating the intensity scale (pulses 196-197).  Note also that 
while much of the ``typical'' core emission just follows the central 
longitude, the strongest such pulses tend to fall nearly on it or even 
precede it---that is, about 1\degr\ earlier.  Both the background noise 
level and interference level of this observation is exceptionally low 
with the latter effectively disappearing into the lowest intensity white 
portion of the $I$ color scale.}
\label{fig2}
\end{center}
\end{figure*} 

\begin{figure} 
\begin{center}
\includegraphics[width=78mm,angle=-90.]{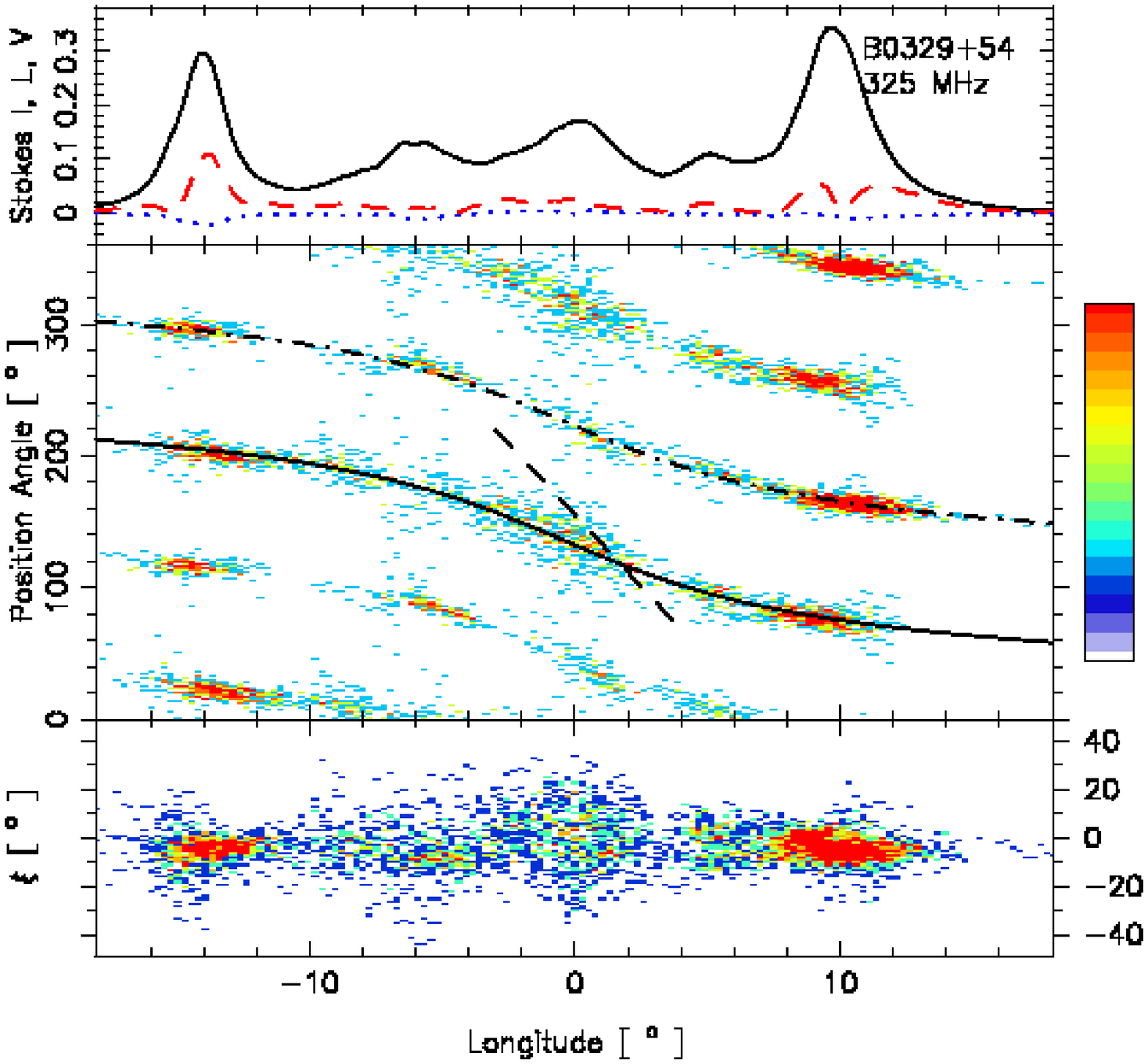}
\caption{Average polarisation and position-angle (PA) distributions  
after Fig.~\ref{fig1} for three intensity sections of the 325-MHz observation 
corresponding to signal-to-noise ratio (hereafter S/N) levels of unity (left), 
2--7 (right top), and 8--12 (right bottom).  Distinguished on a pulse-by-pulse 
basis primarily according to the intensity of the bright central feature, the 
average profile changes from having five clear components at the lowest 
levels to exhibiting little more than the central feature at the highest ones. 
Note that the PA distributions at the lowest intensities accurately follow 
RVM ``tracks''; whereas, those with higher levels of central component 
power exhibit increasingly pronounced non-RVM PA effects.  In particular, 
note that at intermediate intensities (right top) the linear PPM ``kink'' lies 
under the trailing part of the central component just above the PPM RVM 
track, but at the highest levels the ``kink'' departs maximally from the track 
and is centered under the perceptibly shifted central component.  The 
trailing ``spot'' below the PPM track is seen only at the highest intensity 
levels.  The three partial PA distributions represent 305, 1778 and 1191 
pulses, respectively.}
\label{fig3}
\end{center}
\end{figure}

\begin{figure} 
\begin{center}
\includegraphics[width=78mm,angle=-90.]{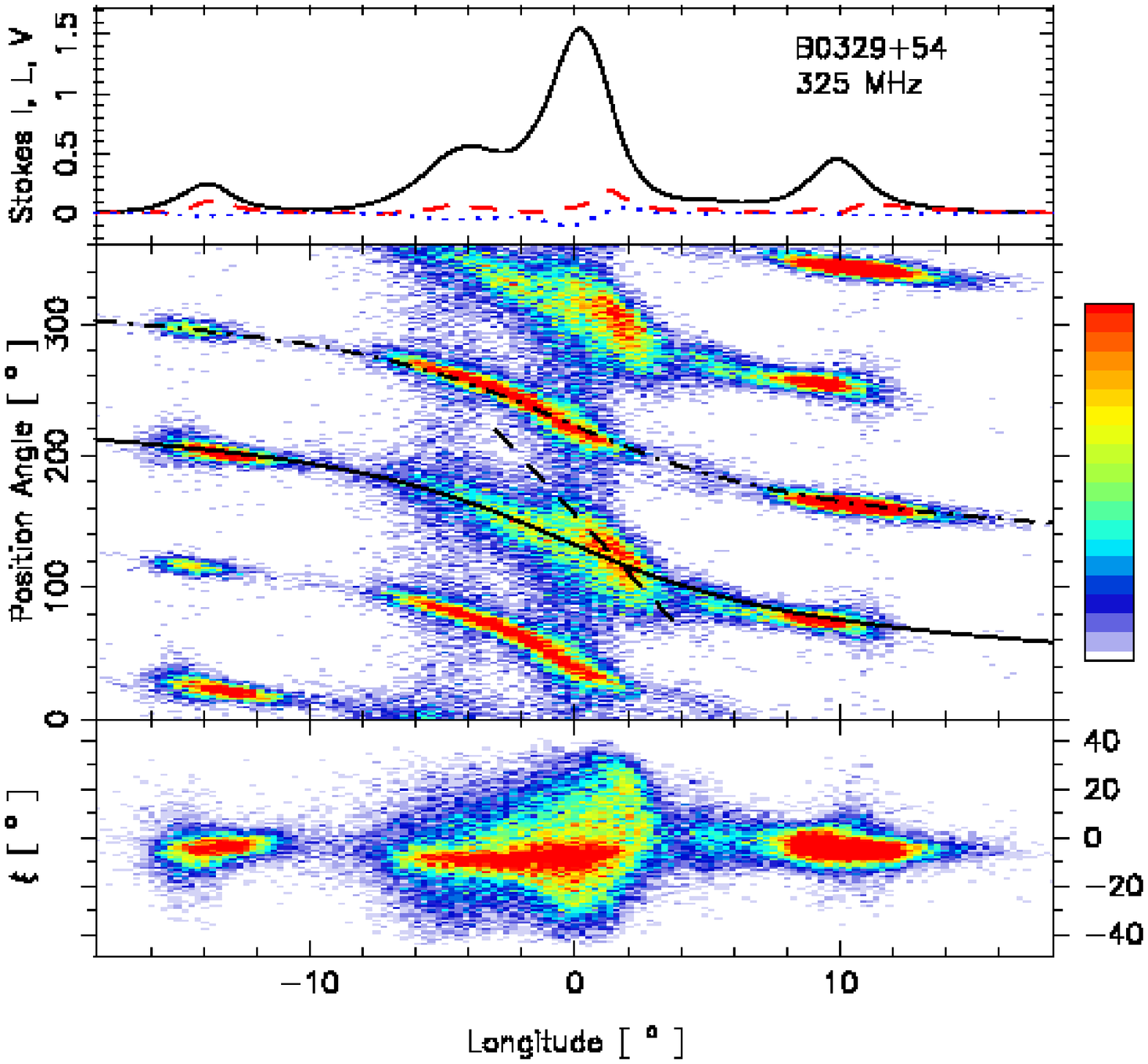}
\includegraphics[width=78mm,angle=-90.]{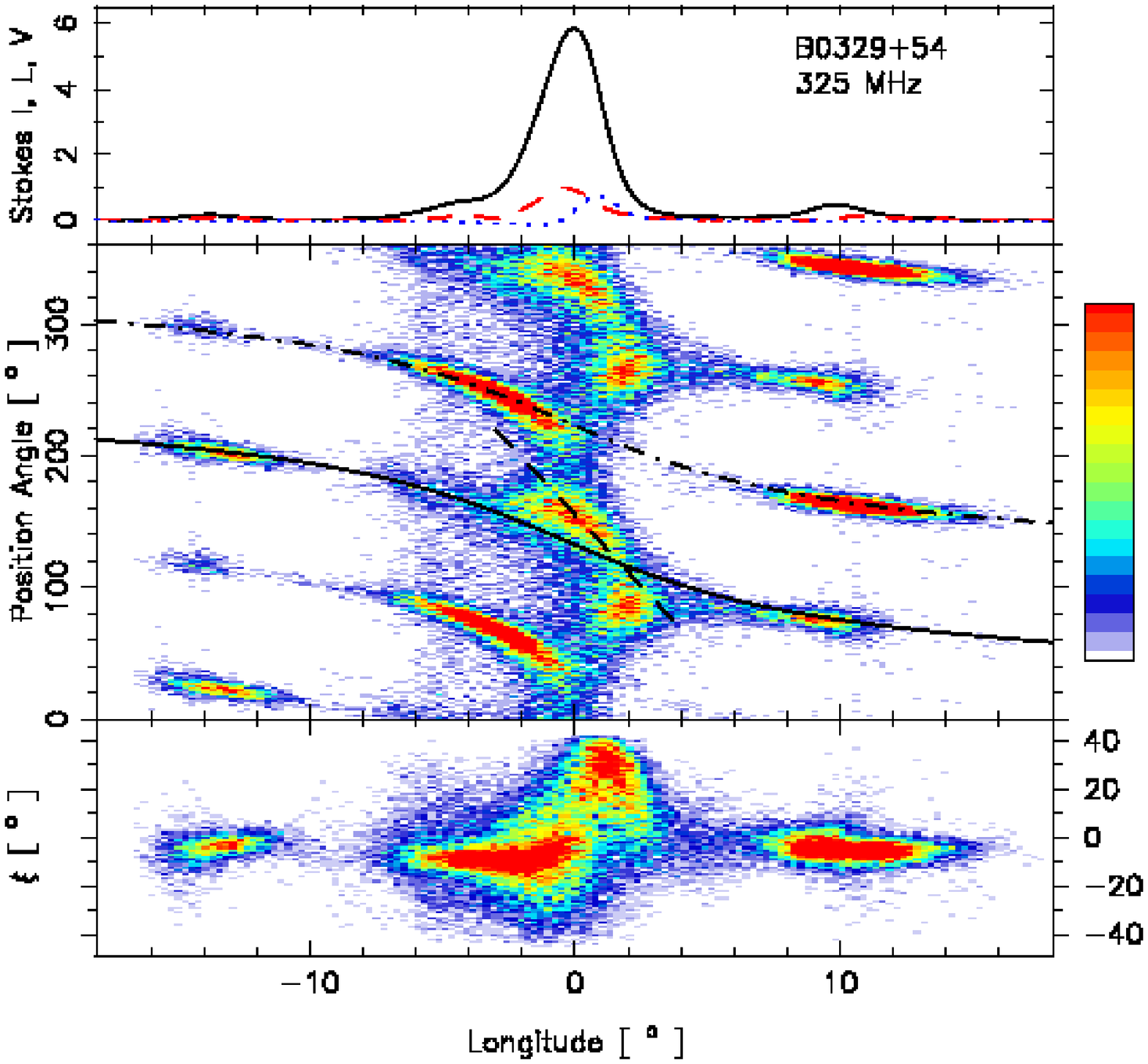}
\end{center}
\end{figure}

\section*{III. Frequency Dependence of Polarization Properties}
\label{sec3}
The three panels of Figure~\ref{fig1} give colour-intensity-coded displays 
of B0329+54's (normal mode) PA occurrence frequency as a function 
of longitude at 325, 610 and 2695 MHz (upper left, bottom and right), 
respectively.  The 180\degr\ PA range is given twice for clarity, and an 
RVM model (discussed below) is overplotted along both the stronger 
``primary'' (solid) and  weaker ``secondary'' (dash-dotted curve) OPM 
(hereafter, PPM and SPM) tracks.  These displays are similar to GL95's 
fig. 1 at 408 MHz but the GMRT observations are more sensitive, better 
resolved and plotted with more intensity levels.  

It is the PPM track, however, which draws the eye:  It too exhibits a 
clear RVM behaviour, but additionally we see a much steeper ``kinky'' 
area under the bright central feature.  This region of conspicuously 
non-RVM dependence within the PA distribution, which has been 
noted in earlier studies as mentioned above, is one of the clearest 
such examples in any pulsar, and we therefore take it as the main subject 
of this investigation.  

The contrasting RVM and non-RVM aspects of B0329+54's PA distribution 
are also clearly seen in ES04's fig. 1.  Their sensitive, well resolved and 
calibrated 328-MHz observation is directly comparable with Fig.~\ref{fig1} 
(left) and shows almost identical non-RVM features.  Both colour-coded 
PA distributions show the RVM behaviour of the SPM track with exceptional 
clarity.  
Many of the same features can also be seen in SP98's 103-MHz PA 
histogram (their fig. 2).  
The signal-to-noise 
ratio (hereafter S/N) and resolution of the observation is inadequate to 
fully distinguish the non-RVM emission; the PPM PAs associated with 
the core component are conflated into an elongated and overexposed 
spot.  However, its elongation on the trailing side to smaller PA values 
appears to be in just the position of the trailing ``spot''.

Remarkably, the 610- (bottom) and 2695-MHz (right) polarisation displays 
show very similar features:   The former is almost identical to the 325-MHz 
behaviour apart from the much weaker leading-component SPM feature.  
The characteristics of the PPM emission, both the RVM track and non-RVM 
``arc'' are virtually indistinguishable.  What subtle differences there are in 
the 610-MHz PA distribution are mostly the results of a somewhat reduced 
S/N.  Even at 2.7-GHz the same features are easily recognizable:  one 
model RVM curve passes through the leading PPM feature, just under the 
steep central ``kink,'' and then again through the trailing PPM conal ``patch''; 
whereas, the SPM curve passes both through the central and trailing features.  
The 2.7-GHz PA distribution thus seems to differ from the others only in subtle 
aspects.  

Focusing on the non-RVM ``kink,'' a thin dashed line (having a slope of 
-21.6 $^{\deg}/^{\deg}$) indicates its position consistently at each 
frequency for convenient comparison.  Apart from a more limited extension 
at 2.7 GHz (which may reflect the poorer S/N), the feature exhibits an 
essentially identical PA-longitude dependence at each frequency.  
The same cannot be said for the trailing ``spot''; its position appears to 
shift slightly between 325 and 610 MHz (from about 85$^{\deg}$ to 
78$^{\deg}$) and then is seen at a dramatically smaller PA at 2.7 GHz 
(about 50$^{\deg}$).

Overall, we find that both the RVM and non-RVM ``kink'' emission features 
in B0329+54's PA distribution have a broadband nature.  We see that the 
position and slope of the linear ``kink'' relative to the PPM track is nearly 
identical at all three frequencies---so that no PA shift is discernible.  Only 
the ``kink'''s extension appears to decrease at the higher frequencies, and 
this may simply be due to decreased S/N. Note also that the trailing-edge 
circular polarization of the core is positive at all frequencies.

\section*{IV.  Intensity Dependence of Polarization Properties}
\label{sec4}

Figure~\ref{fig2} displays a 100-pulse section of the 325-MHz observation 
in Fig.~\ref{fig1} in full polarisation.  Both the strong central feature and 
weak conal outriders are very apparent, and on closer inspection one can 
see that the former exhibits a surprising complexity.  In certain intervals the 
central component is so weak that it seems to be undergoing ``core nulls''; 
at other points (\eg, just before pulse 130) it appears double; and when it 
is emitting most strongly it has a single form that just precedes the longitude 
origin.\footnote{No significant part of these intensity fluctuations can be 
due to interstellar scintillation (ISS) as the decorrelation bandwidth of the 
diffractive ISS is only several 10s of kHz within our 16-MHz bandwidth, and 
the refractive ISS will have a time scale of a few thousand pulses.}  On this basis 
and of course MH93's work, therefore, we began to wonder whether the 
non-RVM portion of the central emission participates in the systematic intensity 
dependence.  

A ready means of exploring this question was that of segregating the 
individual pulses according to their overall intensity relative to the 
off-pulse noise level.  For this we defined a longitude window (--4.13 
to +6.19\degr) that encompassed the bright central component (III) 
under which the non-RVM structure is present.  We used its total 
intensity to determine the mean signal-to-noise ratio (S/N) of each 
pulse within the window and so divide the pulses into respective 
S/N categories. Thereby, the 325-MHz observation was segregated 
into 15 S/N levels. The lowest level was comprised of 197 pulses 
and showed no discernible  non-RVM effects, whereas the higher 
S/N levels exhibited increasingly obvious departures from RVM 
behaviour.  A similar separation was possible for the 610-MHz GMRT 
observation, but at 2695 MHz the overall S/N of the Effelsberg PS 
was poorer so that the lower intensity effects could not be distinguished 
well enough to be useful; it was clear however even here that the 
stronger pulses at this frequency  tended to exhibit the non-RVM 
properties---i.e. the ``kink'' and the ``spot''.

We were consequently able to distinguish the 2970 325-MHz pulses by 
their S/N level over the above range, and each of these intensity ``bins'' 
was plotted in a manner similar to Fig.~\ref{fig1}.  Upon inspection of 
these PA distributions, we found that three distinct kinds of behaviour 
were seen at S/N levels of unity, up to and around 7, and then around 
12---and we then constructed new displays in which the S/N range is 
so fractioned.  Figure~\ref{fig3} (left, right top and bottom, respectively) 
then gives these PA distributions.

The S/N ``binning'' above segregates the PS {\em entirely} on the basis 
of central-component strength.  Thus the three partial profiles show 
clearly that the amplitudes of the leading and trailing conal components 
(I \& IV, as well as the core ``pedestal'') are largely independent of its 
intensity variations.  The very weakest section (left) exhibits a profile 
having five clear components, verifying by a new method the longstanding 
understanding (\eg, Hesse 1973, Kuz'min \& Izvekova 1996) that a weak 
pair of components lies in between the usual four (VI \& V, respectively).  
Here the core ``pedestal'' at about --4\degr\ is almost indistinguishable.  
More pertinent to our present concern, however, is the remarkably ``clean'' 
PA distribution, which clearly defines both the PPM and SPM ``tracks'' 
and shows no hint of non-RVM emission.  

We see a very different configuration at intermediate intensity levels 
(Fig.~\ref{fig3}, top right):  Here the conal outriders have about the 
same amplitudes (as in the earlier low S/N display), but the bright 
central component is some 10 times stronger and is preceded by a 
somewhat weaker ``pedestal'' feature at about --4\degr\ longitude.  
The aggregate linear polarisation in the central profile region is 
slight, but Fig.~\ref{fig2} demonstrates that the fractional linear of 
the individual pulses here often reaches some 50\%---thus the PA 
distributions are very well defined.  It is then completely understandable 
that little aggregate linear survives.  We can discern that the ``pedestal'' 
feature is characterized by strong SPM as well as PPM power that 
exhibits an RVM distribution---and its nearly complete depolarization 
shows that their levels must be about equal.  The non-RVM ``kink'' 
is situated prominently just adjacent to the PPM track, occurring well 
on the trailing side of component III.  Its steeper linear appearance is 
familiar from the distributions of Fig.~\ref{fig1}, and we note that it just 
reaches the PPM track on its trailing end.  It is also worth noting that 
the component III is labeled (here and in some other partial profiles) by 
weak anti-symmetric circular polarisation in a core-like manner---and 
also that the ``kink'' nearly coincides with the positive (lhc) peak.  

Furthermore, the highest intensity pulses (Fig.~\ref{fig3}, bottom right) 
show yet a different behaviour:  Here the bright central component (III) 
is strongest by far---both the ``pedestal'' as well as the conal outriders 
retain nearly equal intensities to those seen at lower S/N levels.  The 
SPM track is very similar to that seen at intermediate intensities, but 
with decreased prominence under both the leading conal component 
and the trailing half of the central component.  The most striking effect 
here, however, is the shift of the non-RVM ``kink'' emission to earlier 
longitudes and the appearance of the trailing ``spot.''  This surely is 
very clear evidence of intensity-dependent behaviour.  

With regard to the ``kink'', the Fig.~\ref{fig3} diagrams do not fully 
exhibit what is seen in the full set of 15 intensity-segregated profiles 
(not shown).  There, at higher levels, the non-RVM PPM emission 
consists of two ``spots'', the later one whose position is fixed and the 
earlier one which shifts progressively earlier as the intensity increases.  
We see this motion of the leading ``spot'' conflated into the elongated 
``kink'' in the mid- and high intensity panels of Fig.~\ref{fig3} (right top 
and bottom).  The ``kink'' shifts are not linear (and the dashed line is 
not a fit, only a reference mark) and seem to curve slightly along the 
PPM track, displaced by perhaps some 1.5\degr; and the 325-MHz
``spot'' lies below the PPM track by a roughly similar amount.  The 
``spot'' is associated with the far trailing edge of the central component, 
which is almost fully left-hand circularly polarized (something also 
very clearly seen in ES04's fig. 1).  This easy to miss circumstance 
now prompts interest in the leading portion of the central component 
where right-handed circular power is often seen, but has disappeared 
in this average of very bright individual pulses.  The dynamics of this 
averaging is very clearly shown in the bottom panel of ES04's fig. 1, 
where the canted band shows how the circular polarization depends 
on longitude---and the leading ``spot'' motion above then links higher 
intensity levels with increasing RH circular polarization (see also 
Fig.\ref{fig6}).  

In summary, Figs.~\ref{fig1}--\ref{fig3} show in different ways that both 
polarisation modes exhibit an RVM behaviour with almost continuous 
tracks at all longitudes---that is, the non-RVM power associated with 
the PPM is seen in addition to a clear RVM track. Only at the highest 
intensity levels do the PA distributions suggest that non-RVM emission 
may displace the RVM emission in certain longitude regions under the 
central component.

\section*{V.  Intensity Dependence of the Bright Central Component (III) and 
its Neighbors} 
\label{sec5} 

Returning now to a fuller study of Fig.~\ref{fig2}, generally, power associated 
with the central component straddles or slightly trails the longitude origin---but 
the most intense subpulses just precede it (just like in MH93).  Additionally, 
``pedestal'' emission can be seen at about --4\degr\ longitude in many pulses, 
often giving the impression that the ``core'' is double.  The fractional linear and circular 
polarisation in this central region is not small, the former typically some 50\% 
(green) but reaching 70\% (yellow).  Right-hand circular predominates prior to 
the longitude origin and left-hand after it, both at typical levels of 40\%.  
Given the slight polarisation of the total profile (see Fig.~\ref{fig1}), it is clear 
that both OPMs are active throughout this central region.  Nonetheless, the 
PA distribution in Fig.~\ref{fig2} indicates that SPM power (magenta) tends 
to predominate on the leading side of the origin and PPM power (cyan) just 
after it, but the pronounced stripes of low linear polarization (blue colour) in 
the second column give some indication of how the depolarisation occurs. 
Further, at pulse phases where the modal power is comparable (\eg, 
about 8.5\degr\ longitude in Fig.~\ref{fig1}), the single pulse fractional 
linear is high in contrast to the low aggregate linear polarization in the 
profile.  This strongly suggests that the OPM's are disjoint (Mckinnon \& 
Stinebring 2000).

 We have segregated the modal power using the three-way method 
described in Deshpande \& Rankin (2001; Appendix), which produces 
PPM, SPM and unpolarised (unPOL) PSs.  Partial profiles corresponding 
to these fractions of the total PS (not shown) indicate that the peak intensity 
of the unPOL profile is about twice that of the PPM profile; moreover, the 
SPM peak is about 40\% that of the PPM peak.  They also show that the 
SPM spans a larger pulse-longitude range than the PPM (see also, GL95: 
fig. 1, SP98: fig.2; and ES: fig. 1).  Overall, then, these results appear 
compatible with the premise that most of B0329+54's depolarised power 
does stem from the incoherent superposition of disjoint PPM and SPM 
radiation. 

\begin{figure} 
\begin{center}
\includegraphics[width=78mm,angle=-90.]{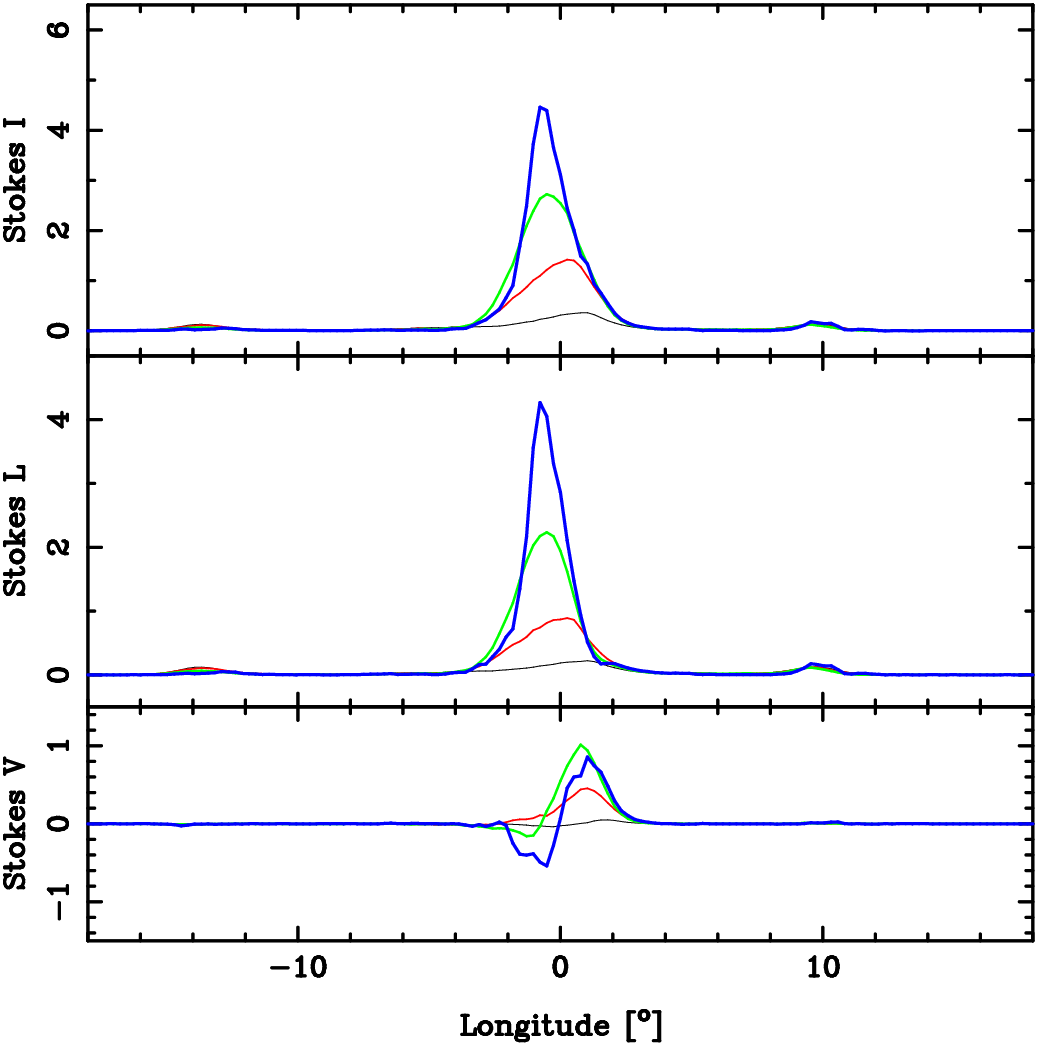}
\caption{Intensity-segregated 325-MHz profiles computed from 
modal PSs segregated using the 3-way segregation technique 
(Deshpande \& Rankin 2001).  Total intensity, linear and circular 
profiles are shown for four intensity levels (S/N levels 1-3, 4-7, 8-11 and 
12-15) in the PPM (left), SPM (right top) and unpolarised (bottom) panels.   
Note the strong intensity dependence of the profiles:  the peak of the central 
component (III) shifts substantially earlier at higher intensities.  The longitude 
origin here is taken at the peak of the 325-MHz total profile as in the earlier 
figures.  Note that the most intense PPM profile has an antisymmetric circular 
signature with a zero-crossing point at the longitude origin, whereas most 
of the emission precedes it.}
\label{fig6}
\end{center}
\end{figure}

Sets of 4 intensity-fractionated profiles are given in Figure~\ref{fig6} for 
the PPM, SPM and unPOL partial PSs, and their three panels depict the 
total power, total linear and circular polarisation.  The PPM (left) central 
component is nearly unimodal, whereas that of the SPM (top right) is, 
as usual, accompanied prominently by the ``pedestal''  feature.  In each 
of the displays the profiles corresponding to the four intensity fractions overlie 
each other, but can readily be identified by their increasing amplitudes.  
All three displays show the intensity-dependent character of component 
III:  this effect is seen strongly in the PPM and unPOL (bottom) plots, but 
is also discernible in the SPM plot.  The PPM total power peak shifts from 
about +0.5\degr\ in the least intense fraction to some --0.5\degr\ at the 
largest intensities, a larger effect by nearly twice than reported by MH93 
for the total profile.  The unPOL profiles also show a dramatic effect, but 
here we see a progressive motion on the leading side of the feature and 
near stasis on the trailing side---such that the high intensity features are 
broader.  The weaker SPM with its ``pedestal'' feature generally peaks 
somewhat earlier and exhibits the least retardation with intensity.  
The most intense PPM power together with the more symmetric unPOL 
power, dominates the total profile form, especially on its leading edge, 
and tends to narrow its overall width.  Indeed, measurement of the 
entire set of 15 intensity-fractionated profiles computed from the total 
PS shows that the trailing half-power point of the central component 
moves earlier by fully 1.5\degr.\footnote{Note that the intensity-dependent 
shift of the core peak is just barely discernible in Fig.~\ref{fig3}, because 
too many intensity levels are conflated.}

\begin{figure} 
\begin{center}
\includegraphics[width=78mm,angle=-90.]{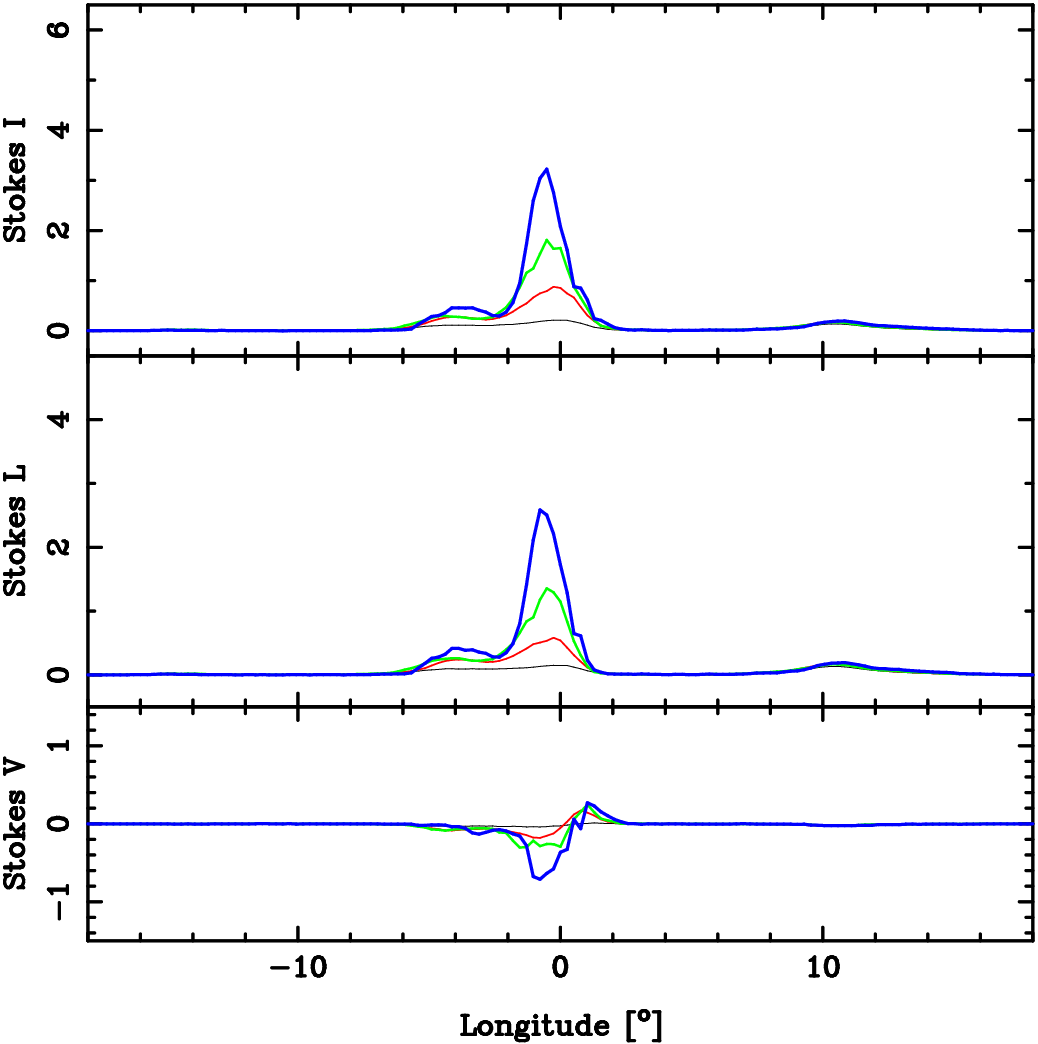}
\includegraphics[width=78mm,angle=-90.]{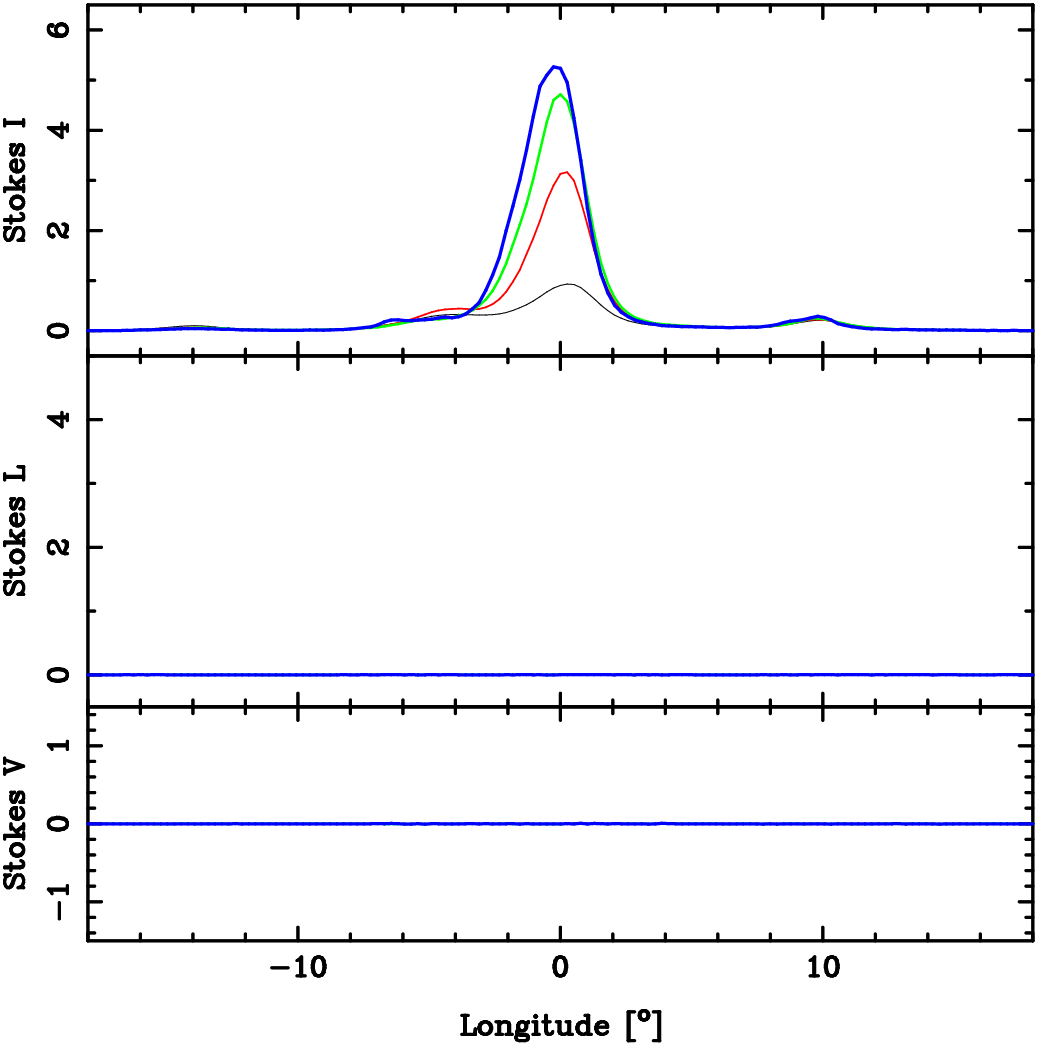}
\end{center}
\end{figure}

We must distinguish carefully between the various contributions to the power 
on the far leading side of the central component.  In the total profiles of 
Fig.~\ref{fig1} this is the region of the ``pedestal'', but in the colour display 
we can see that much of this power consists of a distinct component at some 
--4\degr\ longitude.  Given that individual subpulses can be found at almost 
any longitude in the --5 to --2\degr\ range, one can justly question whether 
this power comprises an actual component, but note that we see the feature 
clearly at intermediate intensities (Fig.~\ref{fig3}, top right) and in the SPM 
profile of Fig.~\ref{fig6} (top right).  Moreover, the nearly 
complete depolarization at this longitude in the total profile indicates that 
PPM power accrues at a similar level and form.  

Conversely, we emphasize that this feature at --4\degr\ longitude contributes 
most of the power to the ``pedestal'' seen in G\&G's fig. 1 as well as our 
Fig.~\ref{fig1} above.  
This component II, however, does not substantially contribute to the total profile 
``pedestal'' (apart, perhaps, from extending its leading edge), and so again 
G\&G's component II and the ``pedestal'' component at --4\degr\ longitude are 
entirely distinct entities. We refer to this feature at --4\degr\ longitude as 
``component X.''

We have computed the auto--correlations (ACFs) and cross--correlations 
(CCFs) of both the natural 325-MHz PS as well as of the 2- and 3-way 
segregated PPM and SPM PSs.  Generally, the ACFs at zero delay show 
a diagonal line of complete correlation as well as symmetrical off-diagonal 
``knots'' of local correlation associated with distinct components---and a 
number are seen, confirming G\&G's finding that the pulsar's emission is 
comprised of multiple components.  These ACFs, however, indicate no 
zero-delay correlation between any of the components, and in particular 
complete independence between fluctuations of the  --4\degr ``pedestal'' 
and central component, even in the SPM.  This surely establishes that 
B0329+54's leading ``pedestal'' is an independent emission feature, and 
we can question whether it is more core- or cone-like, or neither.  Further, 
it indicates that the core's leading ``gouge'' is a part of its structure, not a 
result of ``absorption'' (\eg, Bartel 1981).  CCFs between the two OPM 
PSs at zero delay suggest possible propagation delays between the two 
contributions to core power---again of the order of {$1\over 2$}\degr\ 
longitude---and CCFs at $\pm1$-period delay show significant (typically 
30\%) asymmetric correlation between different parts of the core component.

\section*{VII. Synthetic Analysis \& Discussion} 
\label{sec7}
In the foregoing sections we have reported on our detailed analysis of three 
exceptionally high quality observations of pulsar B0329+54, two made with 
the GMRT at 325 and 610 MHz, and a third from the Effelsberg Telescope at 
2.7 GHz.  Many studies of this pulsar are available in the literature---a number 
of which we have reviewed in the Introduction---but few attempt to delineate 
the properties of the star's central core component.  By contrast, we take this 
central component as our main interest. We discover that the pulsar's polarisation 
properties are strongly dependent on the intensity of the core emission. At 
lower core intensities, the PA behaviour is largely consistent with the RVM, 
which allows us to pursue a model for the pulsar's emission geometry 
(see Appendix A.1).  

\subsection*{Absolute OPM Orientation}
As was well established earlier by BMSH, GL95, Mitra (1999) and 
ES04, the two OPMs can be traced throughout most of B0329+54's 
profile.  Following current practice, we have taken the stronger mode 
as the PPM (\eg, Fig. 1, top left).  However, this delineation of the OPMs 
is no longer adequate.  We require a physical or fundamental geometrical 
basis for distinguishing between the two OPMs, and the well measured 
proper motion of Brisken \etal\ (2002) together with the absolute 
polarimetry of  Morris \etal\ (1979) provides such a means.  

Recently, Johnston \etal\ (2006) have revisited the question of 
whether pulsar rotation axes are aligned with their proper-motion 
directions, making a good case that the natal supernova ``kick'' is 
either parallel or perpendicular to the rotation axis.  Part of this latter 
uncertainty follows from our ignorance about the orientation of a 
specific OPM with respect to the projected magnetic field direction.  
One of us (Rankin 2007) has checked and extended the above work 
and confirms that the case for the star's rotation axis to be fixed 
relative to the supernova-kick direction is indeed very strong. Further 
evidence for such alignment is seen from the x-ray observations of 
young pulsars (Helfand, Gotthelf \& Helpern 2001; Ng \& Romani 
2004).
  
For B0329+54, Morris \etal\ determined that the absolute PA 
(measured ccw from north on the sky) at the peak of the central 
profile component PA$_o$ was 19$\pm$4\degr.  The value, based on the 
observations in Morris \etal\ (1981) at both 1.72 and 2.69 GHz and 
a correct rotation measure, is as accurate as can be obtained with 
average polarization.  We have checked these calculations carefully, 
and used our observations to confirm the relationship between the 
average and RVM ``track'' PAs.
Brisken \etal\ measure B0329+54's proper-motion direction (again 
ccw from north on the sky) PA$_v$ as 119\degr$\pm$1\degr.  The difference 
angle $\Psi$ (=PA$_v$--PA$_o$) is then 100\degr$\pm$4\degr.  
The average PA at the fiducial longitude is distorted by the PPM
``kink'', so Morris \etal's PA$_o$ value probably exceeds the 
RVM PPM by 5-10\degr.  Nonetheless, we find that the pulsar's 
PPM polarization is nearly orthogonal to the projected magnetic 
field direction.  

We reemphasize that the above value applies to the PPM.  If the 
star's natal ``kick'' was delivered parallel to its rotation axis, then 
this identifies the PPM as the OPM orthogonal to the projected 
magnetic field direction.  In their theory of magnetospheric wave 
propagation, Barnard \& Arons (1986) identify the wave polarized 
perpendicular to the projected field direction as the extraordinary 
(X) mode; thus the PPM (solid curve) in Figs.~\ref{fig1} and \ref{fig3} 
can be associated with the X mode and the SPM (dash-dotted)
with the ordinary (O) mode.  

This identification of the SPM with the O propagation mode 
apparently makes this mode subject to refraction; whereas, the 
X mode propagates in a manner independent of refraction. 
Furthermore, the refraction direction is expected to be outward, 
towards the conal edges (Barnard \& Arons 1986; Lyubarskii \& 
Petrova 1997; Weltevrede \etal\ 2005)---that is, outward with 
respect to the magnetic axis---and the PPM/X is the inner of the 
two modes under the main conal components.  Thus, if both 
conal OPMs arise in the same region,\footnote{Our observation 
seems to show good evidence that the two modes are indeed 
arising from the same height. We find that for the mode-segregated 
PPM and SPM, the centers of their outer conal outriders lead the 
steepest gradient point by about 2\degr. By applying the BCW
model this gives an emission height of about 300 km for both the 
modes. However, even better S/N than what we have is needed 
to establish this fact with certainty.} we should expect the 
O-mode emission to have been refracted outward relative to 
that of the X mode.  This behaviour of the outer conal components 
is very usual as we had noted above (Rankin \& Ramachandran 
2003), and differences in the directions of X- and O-mode 
propagation may provide a direct measure of the magnetospheric 
refraction in peripheral regions of a pulsar's polar flux tube.  
Finally, the circumstance that the identified O mode is the outer 
conal emission mode lends support to the premise that 
B0329+54's supernova ``kick'' was indeed aligned with its 
rotation axis.  

Much is yet to be learned about the conal structure of this pulsar 
and its causes.  While two emission cones are well known in 
a number of pulsars, only for B0329+54 is there evidence for 
four (G\&G, see also Appendix~A.1).  In addition to the bright third cone (comps. I \& IV), 
we confirm the innermost cone comprised of G\&G comps. II 
and V and also find that it is PPM (X-mode) dominated (\ie, 
Fig.~\ref{fig3}, left).  Regarding G\&G's second and 
fourth cones, our generally core-directed analyses resulted 
only in the detection of a few weak subpulses at the specified 
longitudes, so insufficient to serve as confirmation.  

\subsection*{A Glimpse of the Polar-Cap Acceleration Process}
Turning now to the perplexing PA ``kink'' in the pulsar's PPM 
traverse, we recall that it a) extends across much of the pulsar's 
core component; b) is associated with a strong intensity-dependent 
shift of the core's position to earlier longitude; and c) is highly 
left-hand (positively) circularly polarised at low intensities, 
but gradually shifts to {--/+} anti-symmetric circular at high 
intensities.  Each of these properties is well demonstrated 
by our analyses above, and overall are reminiscent of KD's 
model for the Vela pulsar.  And we reemphasize that the ``kink'' 
is a frequency-independent feature:  slightly different parts of it 
are revealed at the three frequencies (with their particular S/N 
levels)---and different parts of it clearly correspond to different 
intensity levels---but the feature as a whole exhibits a single 
consistent PA-longitude dependence.

Let us also recall the sightline geometry that the ``kink'' entails.  
On the extreme trailing edge of the core component, the ``kink'' 
reaches and joins the PPM curve, but it extends steeply across 
the longitude origin and departs ever more strongly from the 
RVM track.  Moreover, we know from Appendix~A.1 that 
the sightline impact angle $\beta$ is about --3.5\degr; whereas, 
the half-power edge of the core component may be at some 
(5.5\degr/2 =) 2.8\degr.  Thus the sightline track is poleward of 
the magnetic axis and seemingly just outside the half-power 
intensity level of the magnetic-axis-centered core.  Sightlines 
encountering the core then entail acute values of the magnetic 
azimuth relative to the pole.  And although the O-mode radiation 
does suffer refraction, such refraction would be along radials 
from the magnetic axis and thus would not appreciably alter 
the longitude position of the core emission.

The identification of the PPM ``kink'' with the X mode, as 
demonstrated in the previous section, rules out propagation 
effects as its primary cause.  Other possible grounds dismissing 
propagation effects hinge on the ``kink'''s complete constancy 
over a three-octave frequency band.  We can then understand 
the ``kink'' emission as reflecting A/R effects in an unusual 
viewing or emission geometry (\cf, Blaskiewicz \etal\ 1991).  
We also assume that B0329+54's magnetic field is nearly dipolar.  
Consequently, the ``kink'' indicates emission at increasing 
altitude, mostly along the magnetic axis (with perhaps some 
small displacement to adjacent field lines) as a function of intensity.  
Were the emission occurring at a fixed altitude, we would see 
a different sort of ``kink'', one with a fixed longitude displacement 
from the PPM track.  But in fact, the steeper PA track of the ``kink'' 
indicates that its longitude displacement (from the RVM ``track'') 
increases from zero to some 1.5\degr\ or so over precisely the 
same longitude interval in which the intensity-dependent shift 
of the PPM core emission is observed.  This is just the A/R 
signature of a height-dependent amplification or cascade:  the 
weaker core emission is emitted at low altitude and, as its intensity 
increases, so does its emission height, retarding the total 
intensity and displacing its PA along the ``kink''.  These various 
circumstances are clearly indicated in Figs.~\ref{fig1} \& \ref{fig2}.  

We then have two ways of using the combined effects of A/R to 
estimate the range of emission heights over which this cascade 
or amplification process occurs: we can measure the maximum 
longitude displacement of the ``kink'' from the PPM RVM, or we 
can directly estimate the time interval corresponding to the range 
of height-dependent retardation observed in the core emission.  
Neither is trivial to measure precisely,\footnote{In particular, we 
cannot simply measure the ``kink'' displacement relative to the 
PPM RVM along a particular PA, because these points represent 
different longitudes and thus different sections of the field above 
the polar cap.} but both represent shifts of roughly 1.5\degr\ 
longitude.  The displacements are both produced equally by the 
aberration and retardation, so only half of the shift is associated 
with the retardation and thus the emission-height difference.  
1.5\degr\ then corresponds to 3 ms, and half of this to a height 
difference of some 450 km (Dyks \etal\ 2004).

Finally, let us consider the circular polarization.  Fig.~\ref{fig6} 
indicates that Stokes $V$ is --/+ (rh/lh) antisymmetric 
both in the SPM and in the PPM at high intensity; at low intensity 
the PPM peak falls after the longitude origin and is positively 
circularly polarized.  So, we see that the intense PPM emission 
largely fills the polar cap region and exhibits the antisymmetric 
circular-polarization signature frequently seen in core features.  

Emission associated with the ``kink'' then, appears to provide a 
rare glimpse of the core radiation process.  It reveals ---apparently 
as different pulses sample this process at different phases with  
different intensities and polarizations---what seems 
to be a height-dependent amplification or cascade that moves 
nearly along the magnetic axis, growing ever more intense with 
altitude.  ES and several other authors (Petrova 2006a,b; Melrose 
\etal\ 2006) have tended to interpret the non-RVM emission and 
prominent circular polarization in the B0329+54 core as evidence 
of propagation effects.  
But we now see that the geometrical interpretation can be favoured.
Not only does the phenomenon share some of the features documented 
by KD for the Vela pulsar, but both 
the association of the PPM with the X mode and the broadband 
nature of the ``kink'' rules strongly against propagation.  It would 
seem that ES04 and the others failed to appreciate the primary importance of the 
intensity dependence (MH93) in their interpretations of this pulsar's 
``kink'' radiation.   

One can surely understand how ES04 concluded, from their 
elegant Poincar\'e-sphere analysis, that the non-RVM PPM ``kink'' 
and ``spot'' are propagation effects.  Arguing that the OPM emission 
is initially linearly polarized, and then acquires some circular polarization 
(thus becoming elliptically polarized) in passing through a medium, 
they identify the annulus seen in their 
fig. 2 as evidence of this process.  Such a process, however, requires 
that the OPMs superpose coherently (Mckinnon \& Stinebring 2000), 
or what is the same, they cannot be disjoint---and in B0329+54 there 
is strong evidence that the OPMs {\em are} disjoint:  a) The polarization 
is higher in individual pulses than in their aggregate especially within 
the core region.  b) No strong zero-lag, in-phase correlation was found 
between the PPM and SPM power of the core component (except on its 
far trailing edge).  And c) if O-mode refraction operates as we have 
suggested above, displacing the modal subcones of the principal 
conal components and bending O-mode ``kink'' radiation toward the 
magnetic axis and out of our sightline, then the OPMs are differently 
directed and thus cannot superpose.  In our view, then, the annulus 
is produced by the emission geometry of the intensity-height dependence, 
wherein both the linear and circular polarization exhibit large variations 
(with the circular fully changing sense).

\subsection*{The Perplexing Moving ``Spot''}
In contrast to the ``kink,'' the spot appears to show a strong 
movement with frequency.  Its behaviour can be traced through 
the three panels of Fig.~\ref{fig1}, where we see it centred at 
about 2.2\degr\ longitude, but appearing at 80-90\degr\ PA at 
325-610 MHz and then some 50\degr\ at 2.7 GHz.  The ``spot'' 
is very clearly associated only with the most intense pulses as 
is clear from Fig.~\ref{fig3}.  A less sensitive 21-cm observation 
(not shown) shows a ``track'' of PAs something like that seen 
in the 2.7-GHz panel of Fig.~\ref{fig1} but without the ``spot,'' so 
it is important to further explore this possibly very important 
detailed behaviour.  If a systematic motion of the ``spot'''s PA 
and its dependence with frequency can be verified then it could 
serve as a diagnostic for studying propagation effects in the 
pulsar magnetosphere.  

\subsection*{The Mysterious Component X}
A very important question, which the foregoing analyses 
have pointed strongly (see also Appendix~A.2), is the 
nature of component X.  It is the major part of the ``core 
pedestal'' and peaks very close to --4\degr\ longitude (whereas 
G\&G's component II is at --5.5\degr).  We see in the figures above 
(especially Fig.~\ref{fig6}, top right) that component X is prominent 
in the weaker SPM, but its nearly complete aggregate depolarisation 
(see Fig.~\ref{fig1}) clearly indicates that it must be comprised 
of a comparable amount of PPM power.  Component X appears as 
companion to the core component, but we find no evidence 
for any correlated power between it and the core.  Its intensity 
behaves more like the conal components;  if it participates at 
all in the core's intensity dependence (\eg, Fig.~\ref{fig6}), it does 
so in a much weaker manner.  

The importance of understanding component X is not confined to 
B0329+54 as a number of other pulsars have similar features.  
An excellent example is that of B1859+03 at 21 cms (see 
Radhakrishnan \& Rankin 1990: fig. 1), where we see a fully 
depolarised component prior to the star's core that is prominently 
marked by an antisymmetric circularly polarised signature.  
This leading feature as well as component X appears to be 
almost fully depolarised over its entire width, and this is unlike 
what is seen in outer conal components were the PPM and 
SPM contributions are characteristically displaced somewhat 
in radius and thus in longitude (Rankin \& Ramachandran 
2003).  Petrova (2000) and Weltevrede \etal\ (2005), argue 
that conal emission can be refracted inward due to assumed 
lower plasma densities near the magnetic axis, perhaps 
explaining such a feature; apparently, however, only the 
O mode would be so refracted, so it is difficult to see how 
this could be squared with the prominent depolarization of 
component X.

\subsection*{Concluding Comments}
In the foregoing sections of the paper we focussed on the 
nature of B0329+54's core component and in particular its 
non-RVM PA ``kink'' feature.  We find that this feature, which 
is prominently associated with the PPM, is comprised primarily 
of X-mode radiation---that is, emission whose electric vector 
is oriented perpendicular to the projected magnetic field 
direction. Furthermore the constancy of the``kink'' over a wide 
frequency band rules out the possibility that the effect is due 
to magnetospheric refraction. Rather, we find that the ``kink'' 
must be interpreted in a manner reflecting the geometrical 
exigencies of the emission processes---primarily aberration 
and retardation.  On the other hand, we see what appears to 
be strong evidence for magnetospheric refraction in the outer 
conal components. The association of the O mode with the 
SPM suggests that this mode should be refracted outward, 
and its broader longitude extent appears to bear this out. 
The moving ``spot'' and the X component could also be 
effects of magnetospheric propagation. 

Three key circumstances must be considered in interpreting this 
``kink'' emission:  a) its prominent intensity dependence first noted 
by MH93, and the manner in which the linear polarization of the core 
emission, increasingly retarded with intensity, produces the ``kink''; 
b) the broadband nature of the ``kink'' radiation
and c) the character of its circular polarization, 
which is both broadband in character and apparently geometrical in 
origin.  When considered together, the ``kink'''s properties appear to 
provide a rare glimpse of the core emission process, perhaps a 
cascade or height-dependent amplification.  

Core emission has heretofore been inadequately understood, and 
we hope that these analyses will provide some insights which can 
assist in giving it an improved physical foundation.  Given, we now
know, that ions can be pulled off the surface, it is possible that the 
``kink cascade'' can be interpreted in terms of Cheng \& Ruderman's 
(1980) ion-outflow model.
We note that MH93 followed KD's suggestion in interpreting the 
intensity dependence in terms of this 
model, where the higher intensity of the core could result from increased 
ionic discharge from the neutron star's surface. The retardation could 
then be explained, they argue, as a) an A/R effect due to changes of the 
emission height or b) due to lateral movement of the core-emission region 
as a function of intensity.

While the ion-discharge 
might be the origin of the non-linearity behind the intensity dependence, 
these models provide no further insight into how it is linked to the non-RVM 
linear or circular polarisation. We have considered the possibility
that height dependent A/R effects can give rise to the observed ``kink.'' 
However, other possibilities like the ion flow within the polar flux tube might 
conceivably produce currents which would in turn distort the local magnetic 
field direction and hence the character of the emitted linear and circular 
polarisation.  Overall, it is difficult to 
understand why the non-RVM effects are associated only with a 
single (PPM/X) mode.  If the OPMs were produced by two different charged 
species as proposed by Gangadhara (1995)---then only that one responsible 
for the PPM might vary significantly in order to distort the PA traverse. 
However, these 
and some other suggestions, which were reviewed in the Introduction 
(Gil \& Lyne 1995; Mitra \etal\ 2000; Mitra \& Seiradakis 2004), all 
ostensibly fail to provide as full an explanation as is needed.

\section*{Acknowledgments}
We thank J. Dyks, R.T. Gangadhara, R. Nityananda, S.A. Suleymanova, G.A.W. Wright,
A. Jessner and A. Karasterigou for informative discussions and/or critical readings
of the manuscript and the GMRT operational staff for observing support.
DM would like to thank Alexis 
von Hoensbroech for providing the 2.7-GHz observation used in this 
work and which was also used in Mitra (1999) and  Mitra \etal\ (2000).  We also 
gratefully acknowledge our use of this archival observation from the Effelsberg 
Radio Observatory courtesy of the Max-Planck Institut f\"ur Radioastronomie in 
Bonn. Portions of this work were carried out with support from US NSF Grants 
AST 99-87654 and 00-98685. This work made use of the NASA ADS system.

\bsp

\appendix
\section[]{Emission Geometry}
\subsection*{A.1 Sightline and Conal Beam Geometry}

B0329+54 has a core-cone (Lyne \& Manchester 1988; LM88) 
triple ({\bf T}) profile (R83a), and its core and main conal component 
(I \& IV) pair are seen over most of the about 0.06 to 15-GHz band in 
which it can be detected.  Three main lines of evidence furthermore 
indicate that this pulsar's conal components represent an outer cone: 
a) it has another set of weak components in between the major ones 
[Hesse 1973; Kuz'min \& Izvekova 1996; see also Fig.~\ref{fig3} (left)], 
b) the cone size increases strongly at low frequency (Mitra \& Rankin 
2002; hereafter MR02), and the OPM configuration and depolarisation 
(\eg, Figs.~\ref{fig1}--\ref{fig3}) is typical of outer cones (Rankin \& 
Ramachandran 2003). 

\begin{table}
\label{tab2}
\centering
   \caption{B0329+54 Emission Geometry}
   \begin{tabular}{ccccc}
   \hline
     Method  & $\alpha$ & $\beta$ & Note & Ref  \\
                    & ($\deg$) & ($\deg$) &      &      \\
   \hline
   L\&M  & 30.8  &  2.9   &          & LM88   \\
   $W_{core}$& 32 & ---&          & R90      \\
   ETVI   &   30   &  2.1  &           & R93b$^a$    \\
   PA sweep & 59$\pm$20&--4.5$\pm$2& PPM & GL95  \\
    &42$\pm$20&--3$\pm$2&      SPM &  \\
   This paper--- &  &    & &  \\
   PA sweep & 35.5$\pm$13& --3.7$\pm$1.0&cor 92\%&  this paper\\
   Thorsett & 32.1 & --3.38 & & MR02\\
\hline
\end{tabular}
$^a$This analysis used a steeper average PA sweep rate of --13.5\degr/\degr, 
thus the smaller values of $\beta$.
\end{table}
 
Outer cones are known to have particular dimensions relative to the 
polar cap size (Rankin 1993a,b; hereafter R93a,b).  Its outer cone 
width together with its central PA sweep rate can be used to determine 
both its magnetic latitude $\alpha$ and sightline impact angle $\beta$.  
Several other methods have also been used to estimate these basic 
parameters for B0329+54, including those of LM88, direct PA sweep
fitting (GL95), the core-width method (R90), and a ``Thorsett''-function 
analysis [see DR01, table 2 for a B0943+10 example or Rankin \etal\ 
(2006), table 2 for B0809+74]---and the results of these analyses are 
given in Table~A1.  

\begin{figure} 
\begin{center}
\includegraphics[width=62mm,angle=-90.]{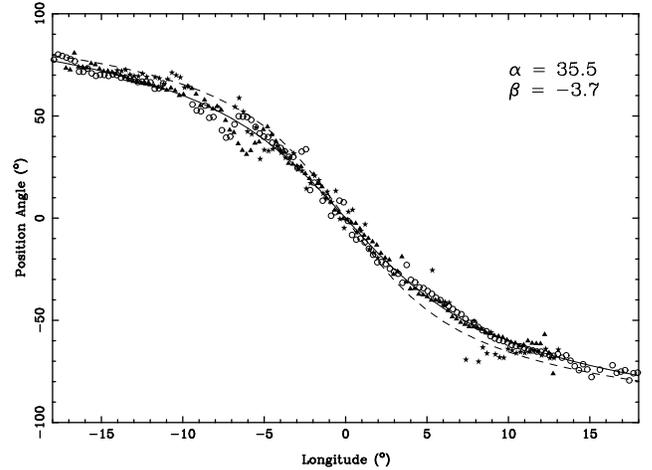}
\caption{RVM fits to the PA traverses.  The 325-MHz data of S/N 1 PPM (filled 
triangle) and SPM (open circle) PA values obtained from  the data of Fig.~\ref{fig3}. 
Each of the tracks were fitted separately and then aligned. The full  
line is the RVM fit to these combined tracks.  The 610-MHz SPM 
(asterik) values pertain to the lowest S/N 1 profile (of 7); a fit was 
used to determine its inflection (steepest gradient) point and then 
to align it with the other curves.  We see here that the PA traverse 
is largely independent of frequency. The dashed line corresponds 
to the average $\alpha$ and $\beta$ values of $51^{\circ}$ and $-4^{\circ}$ 
reported by GL95, which is here clearly seen to deviate from 
our measured values.}
\label{fig5}
\end{center}
\end{figure}

We have carried out our own analyses by fitting the RVM to the PPM and SPM PA traverses 
of the S/N-1 (out of 15) modal PA track at 325- and the SPM PA traverse 
of the S/N-1 (out of 7) 610-MHz observations. 
The low S/N PA tracks were chosen for the fits because  
as seen in Figure~\ref{fig3}, these tracks are not corrupted
by major non-RVM features. 
These three fits yielded mutually consistent $\alpha$ and $\beta$ 
values. The fiducial longitudes obtained for each of these fits were 
used to overlay the three sets of PA values as shown in Figure~\ref{fig5}.  
The combined PAs were then again fitted to eq.(1), and the results 
are summarized in Table~A1. As expected this analysis showed high 
correlations between $\alpha$ and $\beta$, thus the large 
errors---35.5$\pm$13 and --3.7$\pm$1.0\degr, respectively.  The 
fiducial longitude, was well determined at +0.3$\pm$0.5\degr\ 
relative to the peak of the central component in an overall average 
profile.

Table~A1 exhibits that all the available analyses result in similar, and 
generally compatible, estimates for B0329+54's magnetic latitude 
$\alpha$ and sightline-impact angle $\beta$.  The older analyses 
tended to use larger values of the central PA sweep rate $R$, because 
the different OPM behaviours were then unknown.  Larger errors are 
expected from methods which directly fit the PA traverse (and use no 
profile-width information), because $\alpha$ and $\beta$ are highly 
correlated in eq.(1)---as we found above.  Given the coherence of 
these analyses (as well as the weight of the many similar analyses 
applied to many other pulsars which vet and calibrate them), we must 
conclude that B0329+54's magnetic latitude is in the 30--35\degr\ 
range and probably near 32\degr.  Furthermore, its $\beta$ is near 
--3.5\degr, and the negative sense (poleward sightline traverse) is 
indicated by a significantly better goodness of fit.\footnote{ 
In addition to B0329+54's outer conal components (I \& IV) and the 
long known other ones (II, V \& VI), G\&G identified three additional 
components and argued that they represented four concentric cones 
of emission.  They 
then carried out an A/R analysis to compute their emission heights, 
using the central core component as the reference longitude.  Though 
the other inner component pair is too weak to show in the average 
profile, our analysis above does tend to confirm them.  Also, our 
computation of the PA-traverse center above makes it possible to 
apply Blaskiewicz \etal's (1991) method directly.  The results of this 
procedure would, however, provide only a small correction to those 
of G\&G, given that the PA inflection lags the normal-mode core 
peak only slightly.  }

\subsection*{A.2 Core Emission Geometry}
The implication of the foregoing conal analyses is that B0329+54 has an 
angular polar-cap diameter of some 2.45\degr$P_1^{-1/2}$ or 2.9\degr,
where $P_1$ is its rotational period of 0.7145 s and moreover that for 
an $\alpha$ of about 32\degr, we should expect a core-component width 
2.45\degr$P_1^{-1/2}\csc\alpha$ of about 5.5\degr\ longitude.  This line 
of argument has itself been used to determine $\alpha$ for many pulsars 
(R90), and the resulting values in turn found to agree very substantially 
with LM88's method (Bhattacharya \& van den Heuvel 1991). 

Nominally, the core-width estimate has been found to agree best with 
measurements at around 1 GHz as the core components of most objects 
tend to become broader at lower frequencies (Rankin 1983b; hereafter 
B83b).  For B0329+54 measurements show observed normal mode 
core widths in excess of 4\degr\ in the 1--2 GHz band, possibly up from 
some 3.8\degr\ at 15 GHz.  The high quality observations of von Hoensbroech \& 
Xilouris (1997) show a symmetrical, Gaussian-shaped core component 
at 10.55 GHz, but ever more asymmetric ``notched'' forms at longer 
wavelengths: at 4.85 GHz their core is just perceptibly canted on its 
leading edge with a 4.4\degr\ width, whereas at all lower frequencies 
down to our own 610- and 325-MHz observations, the size of the ``gouged'' 
leading region of the core component increases.  This is particularly clear 
in Fig.~\ref{fig1} (right) at 2.7 GHz, where we see a clearly asymmetric 
core component in which a double inflection on its leading edge suggests 
a missing portion.  Also note that the leading ``notch'' is progressively 
deeper at 610 (bottom left), at 325 MHz (top left) and at 103/111 MHz 
(see SP98: fig. 4).  Around 100 MHz the intrinsic core width is some 
3.5\degr\ and corrected values appear to vary little down to 61 MHz 
(Suleymanova 2006).

The strange shape and spectral changes of B0329+54's bright central 
region, we have seen above, have several causes.  First, it is comprised 
not only of the core component (III) but also the ``pedestal'' feature, 
which in turn is comprised mostly of component X but also G\&G's component II.  
Component III, however, surely appears to be the core component.  Its 
antisymmetric circular polarization, when present [\eg, in the SPM, 
Fig.~\ref{fig6} (top right)] tends to have a zero-crossing point near the 
longitude origin taken at the peak of the central component---and we 
have also seen that this aligns closely with the PA inflection point. 
Second, the core-component width falls short of the 5.5\degr\ polar 
cap diameter, thus it does appear to be only partially illuminated.
However, this total profile width is dominated by the bright, narrow, 
earlier-shifting PPM SPs, again contributing to the core's canted 
shape.  It is worth noting that the core's width in the average of the 
weakest pulses (see Fig.~\ref{fig3}, left) could be nearly that of the 
polar cap.  Therefore, we have no basis for regarding the ``pedestal'' 
and component III together as a ``notched'' (or ``absorbed'') core 
component (this line of interpretation was taken in R90).

\label{lastpage}

\end{document}